\documentclass[twocolumn, showpacs, preprintnumbers,amsmath,amssymb]{revtex4}
\usepackage{graphicx}
\usepackage{color}
\usepackage{putexPRL}

\begin{document}

\preprint{PUPT-2308}

\title{Quantum critical superconductors in string theory and M-theory}

\author{Steven S.~Gubser, Silviu S.~Pufu, and Fabio D.~Rocha}

\date{\today}

\affiliation{
Department of Physics, Princeton University, Princeton, NJ 08544, USA
}

\begin{abstract}

We construct zero-temperature solutions of supergravity theories in five and four dimensions which interpolate between two copies of anti-de Sitter space, one of which preserves an abelian gauge symmetry while the other breaks it.  These domain wall solutions can be lifted to solutions of type~IIB string theory and eleven-dimensional supergravity.  They describe quantum critical behavior and emergent relativistic conformal symmetry in a superfluid or superconducting state of a strongly coupled dual gauge theory.  We include computations of frequency-dependent conductivities which exhibit power law scaling in the infrared, with exponents determined by irrelevant perturbations to the symmetry-breaking anti-de Sitter background.

\end{abstract}


\maketitle

\section{Introduction}

In \cite{Gubser:2009qm,Gauntlett:2009dn}, explicit examples of superconducting black holes were exhibited in type~IIB supergravity and M-theory, respectively.  These works follow the general scheme of \cite{Gubser:2008px,Hartnoll:2008vx} for constructing superconducting black holes: a complex scalar charged under an abelian gauged field condenses outside the horizon when the charge of the black hole is big enough.  The constructions of \cite{Gubser:2009qm,Gauntlett:2009dn} draw upon advances including \cite{Corrado:2001nv,Martelli:2004wu,Gauntlett:2009zw,Denef:2009tp} in the understanding of how to embed solutions of gauged supergravity into ten- and eleven-dimensional supergravity.

In \cite{Gubser:2008wz} it was suggested that emergent conformal symmetry should emerge in the zero-temperature limit of superconducting black holes, provided the scalar potential has a symmetry-breaking minimum.  It was further suggested that if there was no such minimum, the zero-temperature limit should involve emergent Lorentz symmetry.  In \cite{Gubser:2008pf}, strong numerical evidence was provided that Lorentzian symmetries do emerge on the thermodynamically favored branch of superconducting black hole solutions to simple theories in $AdS_4$.  It is not hard to produce similar numerical evidence in favor of emergent conformal symmetry when there is a symmetry-breaking minimum in the scalar potential; also, domain wall solutions were constructed in \cite{Gubser:2008wz} which are, plausibly, the zero-temperature limits of the thermodynamically favored superconducting black holes.

In the current Letter, we apply the techniques of \cite{Gubser:2008wz} to the theories discussed in \cite{Gubser:2009qm,Gauntlett:2009dn} to construct domain wall geometries which are candidate ground states for finite-density matter in the gauge theories dual to the $AdS_5$ and $AdS_4$ geometries we consider.  While we would like to go further and claim that the geometries we construct are the genuine ground states of the theories under consideration at finite density, such claims are difficult to establish without knowing the full spectrum of supergravity deformations.

We start in section~\ref{AdS5} with the $AdS_5$ example, and continue in section~\ref{AdS4} with the $AdS_4$ example.
We finish in section~\ref{DISCUSSION} with a brief discussion and a conjecture about the relation between renormalization group flows and emergent conformal symmetry in finite-density systems.

The authors of \cite{Gauntlett:2009dn} anticipated the quantum critical nature of the zero-temperature limit of the superconducting black holes they studied.

\section{A string theory example}
\label{AdS5}
Consider the action
\eqn{Sfive}{
S  = {1\over 2\kappa^2}\int d^5x \sqrt{-g} {\cal L} \,}
with
 \eqn{Lfive}{
  {\cal L} &= R - {1 \over 4} F_{\mu\nu}^2 - 
   {1 \over 2} \left[ (\partial_\mu \eta)^2 + \sinh^2 \eta 
    \left( \partial_\mu \theta - {\sqrt{3} \over L} A_\mu \right)^2
   \right]  \cr
   &\qquad{} + {3 \over L^2} \cosh^2 {\eta \over 2} (5-\cosh\eta) + 
    \hbox{(Chern-Simons)} \,,
 }
where $\eta$ is the magnitude of the complex scalar and $\theta$ is its phase.  The kinetic terms come from the non-linear sigma-model over the Poincar\'e disk, parametrized by $z = e^{i\theta} \tanh {\eta \over 2}$.  The lift of this lagrangian to a class of solutions of type~IIB supergravity, based on D3-branes at the tip of a Calabi-Yau cone, was described in \cite{Gubser:2009qm}.  

The domain wall geometry takes the form
 \eqn{DomainAnsatz}{
  ds^2 = e^{2A(r)} \left[ -h(r) dt^2 + d\vec{x}^2 \right] + 
   {dr^2 \over h(r)} \,,
 }
and has non-zero gauge field $A_\mu dx^\mu=\Phi(r) dt$ and $\eta$.  As in four dimensions \cite{Gubser:2008wz}, any such domain wall supported by matter obeying the null energy condition must have $A$ concave down; and it also follows from the null energy condition that if $h$ is constant in both the infrared ($r \to -\infty$) and the ultraviolet ($r \to +\infty$), then $h_{\rm IR} < h_{\rm UV}$.

The scalar potential in \eno{Lfive} has two extrema, $\eta=0$ and $\eta=\eta_{\rm IR}\equiv\log(2+\sqrt{3})$, and to each of these corresponds an $AdS_5$ extremum of \eno{Lfive} with radius of curvature $L$ and $L_{\rm IR}\equiv 2^{3\over 2} L/ 3$, respectively.  The domain wall solution interpolates between these two $AdS_5$ geometries, similar to the one found in \cite{Girardello:1998pd}.  It differs in that we insist that as $r \to +\infty$ (the ultraviolet),
 \eqn{UVeta}{
\eta \propto e^{ -\Delta_{\eta} A}=e^{-3 A} \,,
}
corresponding to an expectation value for the dimension $3$ operator dual to $\eta$, but no deformation of the CFT lagrangian by it.  Instead, denoting the conserved current dual to $A_\mu$ in the ultraviolet CFT by $J_\mu$, we consider states with finite $\langle J_0 \rangle$ and finite chemical potential $\mu$.  In other words, we add $\mu J_0$ to the CFT lagrangian, which does not by itself break the $U(1)$ symmetry associated with $J_\mu$.  Non-zero $\eta$ does break this symmetry.

We can choose coordinates such that  as $r \to - \infty$
\eqn{IRAdS}{
A \sim {r\over L_{\rm IR}}, \quad h \sim 1, \quad \eta \sim \eta_{\rm IR}, \quad \Phi \sim 0 \,,
}
with exponentially suppressed corrections, which can be obtained from the equations of motion linearized around \eno{IRAdS}. Of particular interest are the first corrections to the scalar and gauge field, 
\eqn{IRetaPhi}{
\eta \approx \eta_{\rm IR} + a_\eta e^{(\Delta_{\rm IR}-4) r/L_{\rm IR}},  \quad \Phi \approx a_\Phi e^{(\Delta_\Phi - 3) r/L_{\rm IR}} \,,
}
where $\Delta_{\rm IR} = 6-\sqrt{6}$ and $\Delta_\Phi=5$.  A formal series solution for $A$, $h$, $\eta$, and $\Phi$ may be developed in the infrared, in powers of $e^{r/L_{\rm IR}}$, with all coefficients determined in terms of $a_\eta$ and $a_\Phi$.

By shifting $r$, we can set $a_\Phi$ to $1$ without loss of generality.  Such a shift adds a constant to $A$, but this constant can be absorbed by rescaling $t$ and $\vec{x}$ by a common factor.  (See \cite{Gubser:2008wz} for a more detailed analysis of a similar case.)  To fix $a_\eta$, one must impose the ultraviolet boundary condition \eno{UVeta}.  There can be several values of $a_\eta$ that satisfy this condition. We will consider the solution for which $\eta$ has the least number of nodes, since this is the solution most likely to be stable. 

\begin{figure}[h]
\begin{center}
 \includegraphics[width=3in]{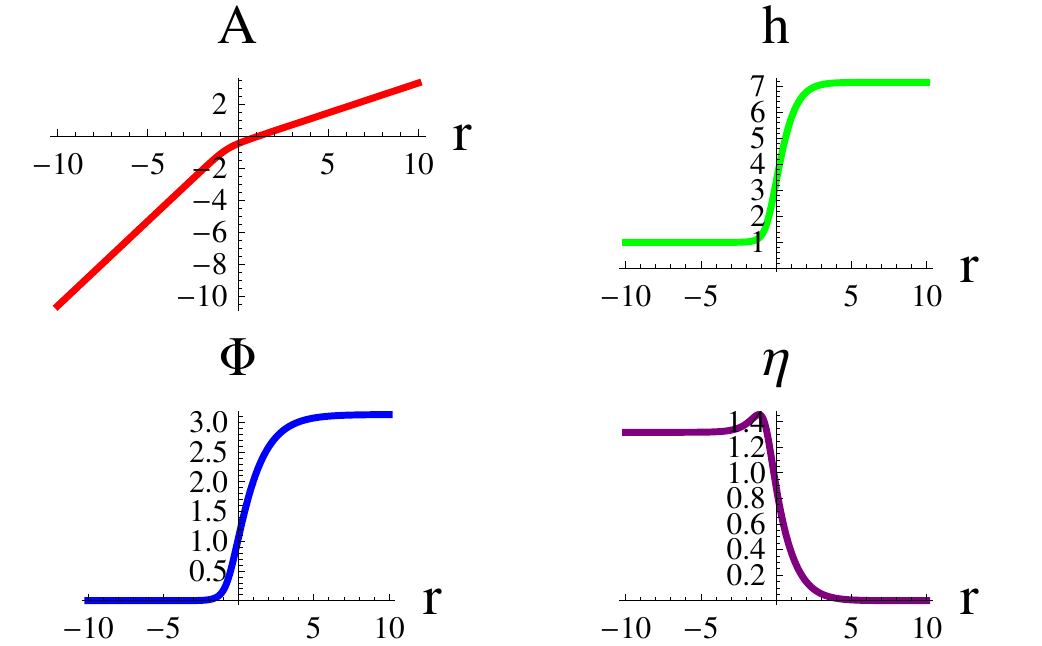}
\end{center}
\caption{
The string theory domain wall.
\label{fig:wallstring}}
\end{figure}

By numerically integrating the equations of motion with the boundary conditions described above we find a nodeless domain wall solution for $a_\eta\approx 2.134$ (see figure~\ref{fig:wallstring}). The relative speed of propagation of lightlike signals in the ultraviolet and the infrared is given by the ``index of refraction" $n\equiv\sqrt{h_{\rm UV}/h_{\rm IR}}$, and for this solution we have $n\approx 2.674$. More complete numerical results are available \cite{numerics}.

One can define a ``normalized'' order parameter through the diffeomorphism invariant formula
 \eqn{OhatDef}{
  \langle \hat {\cal O}_\eta \rangle \equiv 
    \lim_{r \to \infty} {\eta(r) e^{3A(r)} h(r)^{3/2} \over \Phi(r)^3} \,.
 }
In terms of field theory quantities, $\langle \hat {\cal O}_\eta \rangle$ is proportional to ${\langle {\cal O}_\eta \rangle / \mu^3}$, where ${\cal O}_\eta$ is the operator dual to $\eta$ and $\mu$ is the chemical potential.  The proportionality constant depends on the precise normalization one chooses for ${\cal O}_\eta$ and $\mu$.  For our domain wall solution, we find \mbox{$\langle \hat {\cal O}_\eta \rangle \approx 0.322$}. 

It is interesting to note that the ten-dimensional geometry in the far ultraviolet is $AdS_5$ times a Sasaki-Einstein five-manifold ($SE_5$), supported by five-form flux only, whereas in the far-infrared it is of the form first studied in \cite{Romans:1984an}, where a $U(1)$ fiber of the $SE_5$ has been stretched and a combination of the Neveu-Schwarz and Ramond-Ramond two-form gauge potentials have been turned on.  These are not supersymmetric compactifications, so demonstrating stability is non-trivial.

\begin{figure}[h]
\begin{center}
 \includegraphics[width=3in]{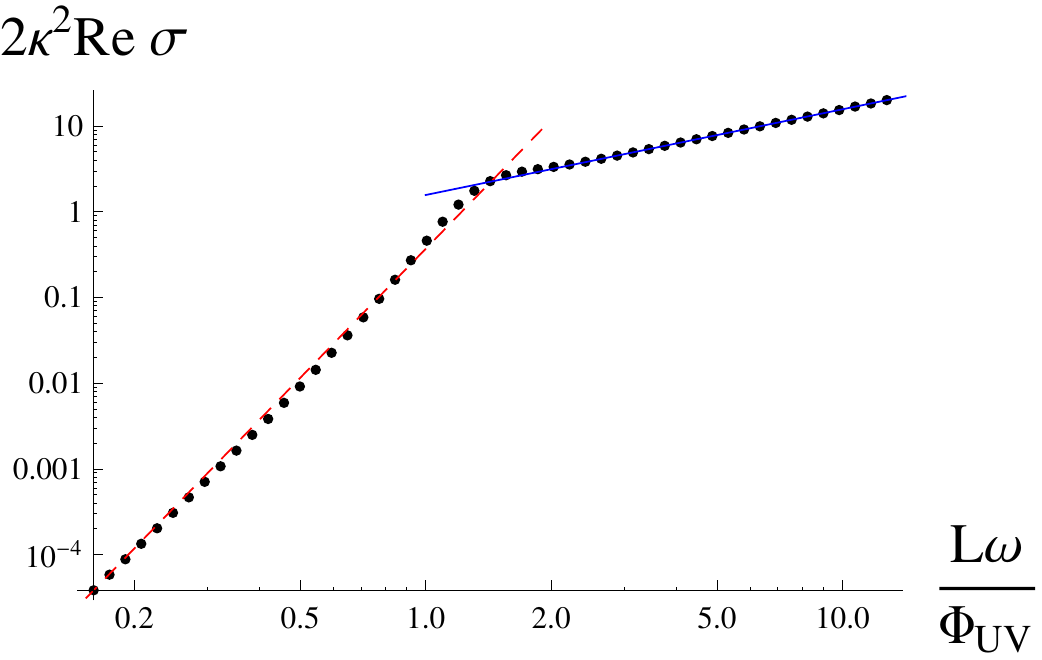}
\end{center}
\caption{
The real part of the conductivity for string theory domain wall. The dots are numerical results,  the dashed line is a $\omega^5$ power law with the coefficient chosen such that the line goes through the first dot in the plot and the solid line is the $AdS_5$ conductivity $\Re\sigma= \pi L\omega/4\kappa^2$.
\label{fig:sigmastring}}
\end{figure}

Having obtained the domain wall solution, we can compute its frequency-dependent conductivity. To this end, we add a time-dependent  perturbation to the gauge field, $A_x = e^{-i\omega t} a_x(r)$ and linearize its equation of motion, obtaining
\eqn{axeom}{
a_x'' + \left(2 A' + \frac{h'}{h}\right)a_x' + \left(\frac{\omega^2 - h \Phi'^2}{e^{2A} h^2} -\frac{3 \sinh^2\eta}{h} \right) a_x=0 \,,
}
with primes denoting $d/dr$. If we solve \eno{axeom}, with infalling boundary conditions in the infrared, the conductivity can then be computed from the ultraviolet behavior of the perturbation. For large $r$,
\eqn{UVax}{
a_x \approx a_x^{(0)} + a_x^{(2)} e^{-2A} + a_x^{(L)} A(r) e^{-2 A} \,.
}
The $A e^{2A}$ term introduces some ambiguity in this computation: it gives a logarithmically divergent contribution to the conductivity \cite{Horowitz:2008bn}. However, since $a_x^{(L)}/a_x^{(2)}$ can be shown to be a real number, this issue only affects the imaginary part of the conductivity, and the real part is unambiguously given by
\eqn{sigmastring}{
\Re \sigma = \frac{1}{2\kappa^2 L} \frac{2 a_x^{(2)}}{i \omega a_x^{(0)}} \frac{h_{\rm UV}}{\Phi_{\rm UV}} \,.
}
Here, the factor of $h_{\rm UV}/\Phi_{\rm UV}$, where $\Phi_{\rm UV}=\Phi(+\infty)$, was introduced to render the conductivity invariant under diffeomorphisms that preserve the form of the metric \eno{DomainAnsatz}. Numerical results for the real part of the conductivity are shown in figure~\ref{fig:sigmastring}. At large frequencies, we recover the $AdS_5$ behavior, $\Re\sigma=L \pi \omega/4\kappa^2$. At low frequencies, we can also obtain the scaling analytically, using the method of matched asymptotic expansions, as in \cite{Gubser:2008wz}. 

The first step is to note that when $r \ll -L_{\rm IR}$, the corrections to \eno{IRAdS} are suppressed.  When they are ignored, \eno{axeom} can be solved analytically.  The infalling solution is
 \eqn{axIR}{
  a_x^{\rm IR} = e^{-r/L_{\rm IR}} H^{(1)}_{\Delta_\Phi -2}\left(\omega L_{\rm IR} e^{-\frac{r}{L_{\rm IR}}}\right) \,,
 }
where $H^{(1)}$ is a Hankel function.  The next step is to note that when $r \gg L_{\rm IR} \log \omega L_{\rm IR}$, one may drop $\omega$ from \eno{axeom} altogether.  The resulting equation probably can't be solved analytically, but the point is that the solutions to \eno{axeom} which determine the conductivity don't depend on $\omega$ in the region $r \gg L_{\rm IR} \log \omega L_{\rm IR}$, except for an overall multiplicative factor: they are given simply by the zero frequency solution.  Provided $\omega L_{\rm IR} \ll 1$, there exists a window $L_{\rm IR} \log \omega L_{\rm IR} \ll r \ll -L_{\rm IR}$ where \eno{axIR} may be matched onto the zero-frequency solution.  The result of this matching is that $a_x^{(0)} \sim \omega^{-\Delta_\Phi+2}$.  To extract the real part of the conductivity, first define
\eqn{current}{
{\cal F}=\frac{h e^{2A}}{2i}a_x\overleftrightarrow\partial_r a^*_x}
and note that ${\cal F}$ is independent of $r$. Inserting \eno{UVax} into \eno{current} it follows that ${\cal F}= \sigma \Phi_{\rm UV}\omega \big| a_x^{(0)}\big|^2$.  On the other hand, inserting \eno{axIR} into \eno{current} shows that ${\cal F}$ is $\omega$-independent.  So we find that 
\eqn{sigmalow}{
\Re \sigma \propto  \omega^{2\Delta_\Phi-5} = \omega^5 \,,
}
where in the last step we used $\Delta_\Phi=5$.  The result $\Re\sigma \propto \omega^{2\Delta_\Phi-5}$ is clearly more general: it basically depends on having good control over the series expansion of the background in the infrared.

As figure~\ref{fig:sigmastring} shows, numerical evaluations of the conductivity interpolate quite smoothly between the infrared and ultraviolet limits just discussed.

\section{An M-theory example}
\label{AdS4}

The four-dimensional theory
 \eqn{Lfour}{
  {\cal L} &= R - {1 \over 4} F_{\mu\nu}^2 - 
   {1 \over 2} \left[ (\partial_\mu \eta)^2 + \sinh^2 \eta 
    \left( \partial_\mu \theta - {1 \over L} A_\mu \right)^2 \right]  \cr
    &\qquad{} + {1 \over L^2} \cosh^2 {\eta \over 2} (7 - \cosh\eta)
 }
derived as a consistent truncation of M-theory in \cite{Gauntlett:2009dn, Gauntlett:2009zw}, clearly is nearly identical to \eno{Lfive} \footnote{The notation of \cite{Gauntlett:2009dn, Gauntlett:2009zw} is related to ours by $\hat A_1=A$ and $\hat\chi=\sqrt{2}  e^{i\theta}\tanh {\eta\over 2}$.}.  As mentioned in \cite{Gauntlett:2009dn}, this truncation is consistent only when $F \wedge F = 0$.

This theory also admits a domain wall solution.  The asymptotically $AdS_4$ geometry is of the same form as \eno{DomainAnsatz}, and now the two extrema of the potential are at $\eta=0$ and $\eta_{\rm IR}=\log(3+2^{3/2})$, corresponding to $AdS_4$ solutions with radii of curvature $L$ and $L_{\rm IR}\equiv \sqrt{3} L/2$, respectively.  If we assume the scalar goes to the second fixed point in the infrared, then as $r\to -\infty$, 
\eqn{IRetaPhiMTheory}{
\eta \approx \eta_{\rm IR} + a_\eta e^{(\Delta_{\rm IR}-3) r/L_{\rm IR}},  \quad \Phi \approx  e^{(\Delta_\Phi - 2)r/L_{\rm IR}} \,,
}
with $\Delta_{\rm IR}=(3+\sqrt{33})/2$ and $\Delta_\Phi=4$.  Imposing no symmetry-breaking deformation of the UV CFT means demanding that
\eqn{UVetaMTheory}{
\eta \propto e^{-\Delta_{\eta} A }=e^{-2A} \,.
}
We numerically found a solution for $a_\eta\approx 1.256$ with index of refraction $n\approx 3.775$.  The normalized order parameter analogous to \eqref{OhatDef} is in this case given by the diffeomorphism invariant formula
 \eqn{hatODefMThy}{
  \langle \hat {\cal O}_\eta \rangle \equiv 
    \lim_{r \to \infty} {\eta(r) e^{2A(r)} h(r) \over \Phi(r)^2} \,,
 }
and is proportional to $\langle {\cal O}_\eta \rangle / \mu^2$.  Our domain wall solution has $\langle \hat {\cal O}_\eta \rangle \approx 0.201$.  Again, more complete numerical results are available \cite{numerics}.

As already noted in \cite{Gauntlett:2009dn}, the infrared geometry in eleven dimensions is an $AdS_4$ compactification of the form studied in \cite{Pope:1984bd,Pope:1984jj}.  The whole geometry is non-supersymmetric, so it is difficult to definitely establish stability.

\begin{figure}[h]
\begin{center}
 \includegraphics[width=3in]{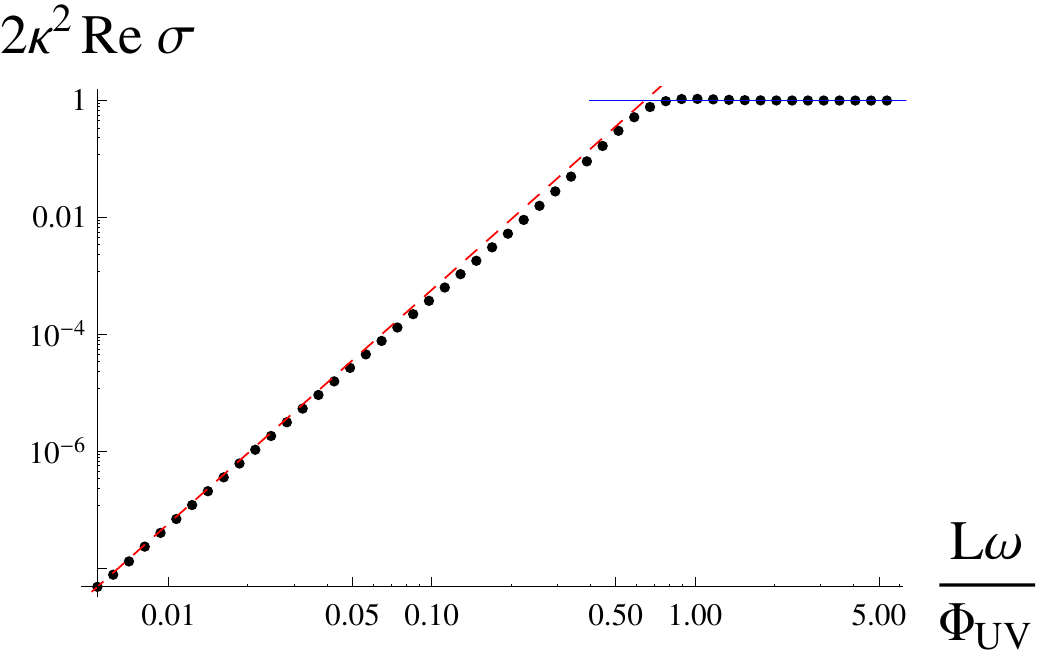}
\end{center}
\caption{
The real part of the conductivity for M-theory domain wall. The dots are numerical results,  the dashed line is a $\omega^4$ power law with the coefficient chosen such that the line goes through the first dot in the plot and the solid line is the $AdS_4$ conductivity $\sigma= 1/2\kappa^2$.
\label{fig:sigmamtheory}}
\end{figure}

The computation of the conductivity is similar to before, so we will be brief.  The main difference is that the behavior of the solutions as $r\to+\infty$ is
\eqn{UVaxMTheory}{
a_x \approx a_x^{(0)} + a_x^{(1)} e^{-A} \,,
}
and this time there is no ambiguity in the imaginary part, the conductivity being given by
\eqn{sigmamtheory}{
\sigma = \frac{1}{2\kappa^2 L}\frac{a_x^{(1)}}{i \omega a_x^{(0)}} \sqrt{h_{\rm UV}}\,.}
Numerical results are shown in \ref{fig:sigmamtheory}. For high frequencies, the conductivity asymptotes to the $AdS_4$ value $\sigma=1/2\kappa^2$ \cite{Herzog:2007ij} and for low frequencies the behavior can be determined analytically with an argument similar to the one described in the previous section. As was shown in \cite{Gubser:2008wz}, in $AdS_4$ the scaling is $\Re \sigma \propto \omega^{2\Delta_\Phi -4}=\omega^4$, and this agrees with the numerical results \footnote{In \cite{Gubser:2008wz}, a $\Delta_\Phi$ was shifted by one unit relative to the definition used here. This explains the apparent discrepancy in the power law.}.

\section{Discussion}
\label{DISCUSSION}

The domain walls we have constructed can be fairly described as superconductors because they spontaneously break the $U(1)$ gauge symmetry associated with the field strength $F_{\mu\nu}$ in \eno{Sfive} and \eno{Lfour}.  According to the arguments of \cite{Weinberg:1986cq}, much of the basic phenomenology of superconductors, including infinite DC conductivity, follows from this spontaneous symmetry breaking.  The domain walls can fairly be characterized as quantum critical because relativistic conformal symmetry emerges in the infrared, and observables, in particular $\Re\sigma(\omega)$, have power-law scaling in the infrared with non-trivial exponents.

Clearly, the domain walls we have constructed are close relatives to holographic renormalization group flows from one conformal field theory to another.  The main qualitative difference is that the breaking of the $U(1)$ symmetry was soft in the RG flows, whereas it is spontaneous in our domain walls.  More explicitly: the RG flows are triggered by adding a relevant operator, dual to the scalar $\eta$ in both cases, which breaks the $U(1)$ symmetry; our domain walls, on the other hand, have by design no such relevant deformation, but instead a spontaneously generated expectation value of the symmetry breaking operator.

It is natural to ask how general the relation between renormalization group flows and emergent conformal symmetry of finite-density matter might be.  Here is a conjecture which makes sense to us:
 \begin{itemize}
  \item Assume that a field theory is well-defined in the ultraviolet and possesses a continuous symmetry.  This ultraviolet theory need not be conformal.
  \item Assume also that if the ultraviolet theory is appropriately deformed, a renormalization group flow results whose infrared limit is a fixed point which breaks the continuous symmetry.
 \end{itemize}
Then the conclusion is:
 \begin{itemize}
  \item The ultraviolet theory, or some deformation of it by operators which do {\it not} break the continuous symmetry, has a finite density, zero temperature state whose infrared behavior is governed by the same infrared fixed point.  Finite density means that the time component(s) of the Noether current(s) associated with the continuous symmetry have finite expectation values.
 \end{itemize}
We are aware of one way to break this conjecture \cite{GubserNelloreForth}: it can happen that the conserved current of the ultraviolet theory flows to a relevant operator at the infrared fixed point.  When that happens, it's impossible (or at least fine-tuned) for the dynamics of finite-density matter to flow to the fixed point.  What we suggest as a real possibility is that relevance of current operators with expectation values in the finite-density state is the only obstacle to the conjecture as we have phrased it.  Since the idea is to systematically pair an RG flow to an infrared critical point with quantum critical behavior of a finite-density state, let us refer to our suggestion as the ``Criticality Pairing Conjecture,'' or CPC.

When applied to the gauge-string duality, the CPC implies the existence of a number of domain wall solutions interpolating among critical points of the scalar potential of gauged supergravity theories.  The CPC might also be tested in situations where some non-string-theoretic approximation scheme can be found, like a large $N$ expansion with perturbative control; or, perhaps, it could be investigated in the context of rational conformal field theories in $1+1$ dimensions.

\section*{Acknowledgments}

We thank C.~Herzog, A.~Nellore, T.~Tesileanu for discussions and for collaboration on closely related topics. We also thank A.~Polyakov, S.~Sondhi, and A.~Yarom for useful discussions.  This work was supported in part in part by the DOE under Grant No.\ DE-FG02-91ER40671 and by the NSF under award number PHY-0652782.  FDR was supported in part by the FCT grant SFRH/BD/30374/2006.

\bibliographystyle{ssg}
\bibliography{embed}

\end{document}